\documentclass[fleqn,10pt]{wlscirep}
\usepackage{graphicx}
\usepackage{subfigure}

\begin{document}
\title{Conformal Invariance of Graphene Sheets}

\author[1,*]{I.Giordanelli}
\author[1]{N. Pos\'e}
\author[1]{M. Mendoza}
\author[1,2]{H. J. Herrmann}

\affil[1]{ETH
  Z\"urich, Computational Physics for Engineering Materials, Institute
  for Building Materials, Wolfgang-Pauli-Strasse 27, HIT, CH-8093 Z\"urich, Switzerland}
\affil[2]{Universidade
  Federal do Cear\'a, Departamento de F\'isica, Campus do Pici, 60455-760 Fortaleza, Cear\'a, Brazil}

\affil[*]{gilario@ethz.ch}

\date{\today}
\begin{abstract}
Suspended graphene sheets exhibit correlated random deformations that can be studied under the framework of rough surfaces with a Hurst (roughness) exponent $0.72 \pm 0.01$. Here, we show that, independent of the temperature, the iso-height lines at the percolation threshold have a well-defined fractal dimension and are conformally invariant, sharing the same statistical properties as Schramm-Loewner evolution (SLE$_{\kappa}$) curves with $\kappa=2.24\pm0.07$.
Interestingly, iso-height lines of other rough surfaces are not necessarily conformally invariant even if they have the same Hurst exponent, e.g. random Gaussian surfaces. We have found that the distribution of the modulus of the Fourier coefficients plays an important role on this property. 
Our results not only introduce a new universality class and place the study of suspended graphene membranes within the theory of critical phenomena, but also provide hints on the long-standing question about the origin of conformal invariance in iso-height lines of rough surfaces.  

\end{abstract}

\maketitle

Rough surfaces are very common in nature and can be found, for instance, in real landscapes \cite{coastlines} and growth surface processes \cite{PhysRevLett.100.044504}. In many cases, they can be characterised by a Hurst (roughness) exponent that describes the height-height correlations of the surface, and consequently, being called self-affine. Random Gaussian surfaces (RGS) with positive Hurst exponents are examples of rough self-affine surfaces, and their use has become very popular since they are analytically tractable. Recently, it was suggested that iso-height lines in this type of RGS are not conformally invariant \cite{RGS_SLE}, since their statistics is not compatible with the Schramm-Loewner Evolution (SLE) theory \cite{Schramm00,  Schramm00, Lawler04, Schramm09, Daryaei12, Pose14} (random curves satisfying SLE statistics are necessarily conformally invariant). However, the fact that iso-height lines of other self-affine rough surfaces, e.g. some grown surfaces \cite{Saberi10, Saberi08}, follow SLE theory opens the question on which are the main properties that are responsible for conformal invariance.

\begin{figure}
\begin{center}
\includegraphics [width=0.5\columnwidth]{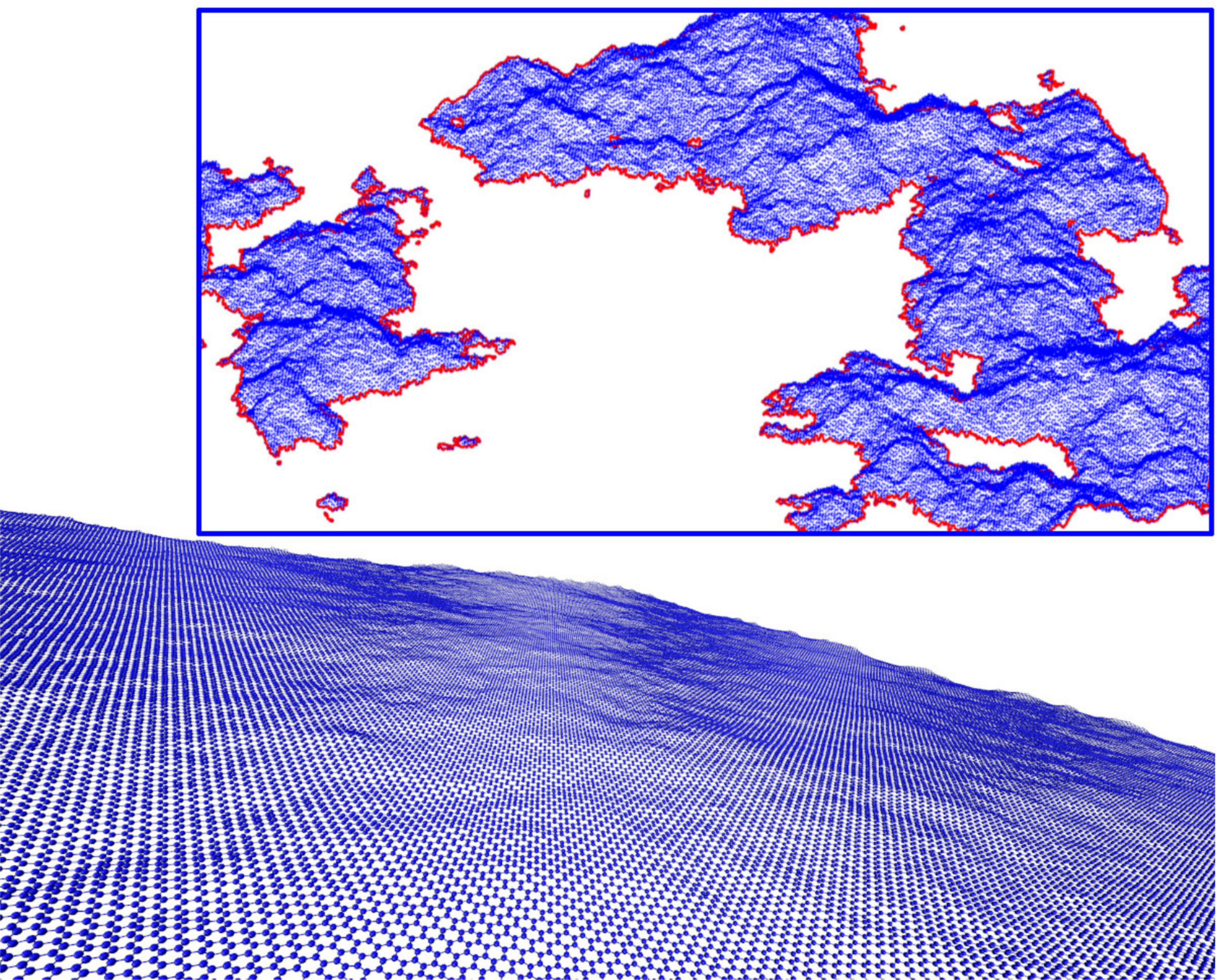}
\caption{Graphene membrane after thermalisation. Inset: The blue points represent carbon atoms that are above the iso-height plane. The red line shows the contour lines, i.e. the intersection between the membrane and this iso-height plane.}
\label{fig:snapshot}
\end{center}
\end{figure}

Graphene, consisting of literally a single carbon monolayer, represents the first instance of a truly two-dimensional material (see Fig.~\ref{fig:snapshot}) \cite{PhysToday,natletter,Novoselov22102004}. It owes its stability to the anharmonic coupling between bending and stretching modes, and the resulting deformation in the third dimension \cite{intrinsic-ripples}.
The study of these deformations, often called ripples, is very important because they affect the electronic and mechanical properties.
For instance, it has been shown that reducing height fluctuations in graphene samples increases their electronic mobility and electrical conductivity \cite{0953-8984-26-13-135303}. 
Since previous numerical studies have shown that the structure of graphene possesses self-affine properties \cite{RoughnessExponent},
in this report, we show that iso-height lines, extracted at the percolation threshold, have a well-defined fractal dimension and are conformally invariant (note that to obtain the percolation threshold one starts from the highest point of the membrane and systematically lowers down the height until a cluster, formed by carbon atoms bounded by covalent bonds to their neighbours, meets both opposite sides of the membrane. This resulting height-threshold is the percolation threshold). Surprisingly, the observed statistical properties do not depend on the temperature of the samples. Furthermore, by exploring the main differences between graphene sheets and self-affine RGS with the same Hurst exponent, we have found that, even if they both have random uncorrelated phases in Fourier space, the distributions of the modulus of the Fourier coefficients are different. Therefore, one can conclude that this distribution is important for conformal invariance, and opens a new way to construct self-affine surfaces that satisfy SLE statistics. Finally, the fact that the study of graphene can be placed within the theory of critical phenomena suggests that the interplay between thermal fluctuations and anharmonic coupling between bending and stretching modes induces a certain degree of criticality in graphene.

In this study, free standing graphene membranes are simulated using molecular dynamics with the adaptive intermolecular reactive bond-order (AIREBO) potential \cite{stuart2000reactive},
which can describe graphene membranes accurately \cite{zhao2009size}. 
 We have imposed free boundary conditions and restricted the analysis of $L \times L$ graphene membranes to the inner area $\big[ \frac{L}{4},\frac{3L}{4}  \big] \times \big[ \frac{L}{4},\frac{3L}{4}  \big]$. 
 
We have performed simulations of quadratic graphene membranes of length $200$~\AA, $400$~\AA, and $800$~\AA, at different temperatures, i.e. $100$~K, $300$~K, and $600$~K. 
A simulation time step of one fs was enough to capture the carbon-carbon interactions. We let the samples evolve $1.4$~ns, $3$~ns, and $13$~ns, for system sizes $800$~\AA, $400$~\AA, and $200$~\AA, respectively. We control the temperature with a Nos\'{e}-Hoover thermostat. Each simulation starts with a flat graphene membrane located in the $x-y$ plane having  small random perturbations in $z$-direction (to break the symmetry) and zero initial velocity (zero temperature). For each temperature and system size, we perform three independent simulations with different initial perturbations. 
We use an equilibration time of $0.2$ ns to reach the desired temperature and verify that the respective Maxwell-Boltzmann distribution for the velocities is recovered (see Fig.~S1 of the supplementary material). From each simulation, we extract graphene sheets in intervals of $5$~ps.   
Following this procedure, we obtain for each temperature up to $720$, $1680$, and $7680$ graphene sheets for $800$~\AA, $400$~\AA, and $200$~\AA, respectively.

In order to compare graphene membranes with RGS, we generate the latter  ones by using the so-called Fourier Filtering Method (FFM). In contrary to graphene sheets, where simulations are performed in real space expressing physical interactions between atoms, RGS are constructed in Fourier space, and can be designed to have a desired Hurst exponent. In the FFM method, the Fourier coefficients of RGS are constructed using the decomposition $u_{\vec{q}}=\sqrt{S(\lvert \vec{q} \rvert )}\tilde{u}_{\vec{q}}e^{i\phi_{\vec{q}}}$, where $\tilde{u}_{\vec{q}}$ are standard Gaussian random variables, the phase $\phi_{\vec{q}}$ are uniformly random variables, and $S(q)$ determines the Hurst exponent $\alpha$ via the relation $S(q)\sim \lvert \vec{q} \rvert ^{-2(\alpha+1)}$ \cite{WKt}. 
Thus, by calculating the inverse Fourier transform of $u_{\vec{q}}$, one gets the RGS. 
\begin{figure}
\begin{center}
\includegraphics [width=0.6\columnwidth]{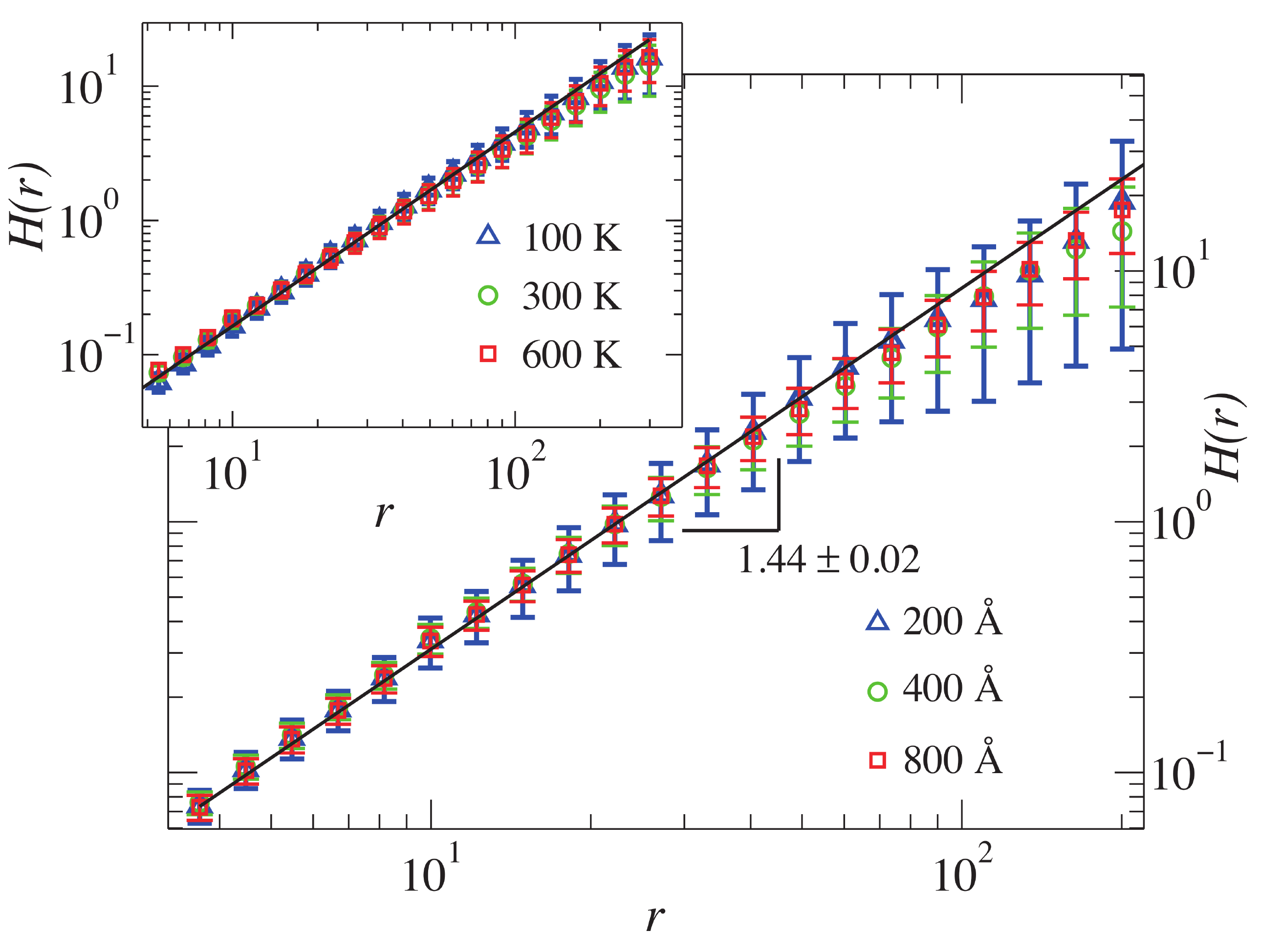}
\caption{The height-height correlation function of the carbon atoms of the graphene membrane.  Main panel: For different system sizes at $T=300$ K. Inset: For different temperatures and a system size of $800$~\AA.}
\label{fig:Hurst}
\end{center}
\end{figure}

Using the thermalized graphene samples, we first compute the height-height correlation function, defined by $H(\vec{r}) =  \langle \vert h(\vec{x}+\vec{r})-h(\vec{x})\vert ^ 2 \rangle$, where $h$ is the height profile of the surface defined by the $z$ component of the atoms. For self-affine rough surfaces, the function $H(\vec{r})$ exhibits power-law behaviour, i.e. $H(\vert \vec{r} \vert) \propto \vert \vec{r} \vert^{2\alpha}$, where the Hurst exponent $\alpha$ characterises the roughness of the surface. We take the statistical averages ($\langle ... \rangle$) considering only carbon atoms that have a certain distance from the boundary to avoid boundary effects. The results presented in Fig.~\ref{fig:Hurst} indicate that within error bars the obtained Hurst exponent $\alpha =0.72 \pm 0.01$ is independent of both, system size and temperature, and is in agreement with the previous work by Abedpour \textit{et al.} \cite{RoughnessExponent}, where it was found $0.6 < \alpha < 0.74$, even though they used different boundary conditions. 
The existence of a Hurst exponent confirms the self-affinity property of graphene membranes and thus sets our graphene sheets in the context of rough surfaces (the measurement of the Hurst exponent using the Fourier space can be found in Fig.~S2 of the supplementary material). Furthermore, knowing $\alpha$ allows us to construct RGS with the same Hurst exponent.

Contour lines of self-affine surfaces exhibit scale invariance, as it is also the case for RGS \cite{PhysRevLett.74.4580}. To verify this property, we extract the contour lines for each graphene sheet (see Fig.~\ref{fig:snapshot}) and each RGS. In the case of graphene (RGS), we consider atoms (grid points) of the surface that are at the percolation threshold and construct the contour lines.
\begin{figure}
\begin{center}
\includegraphics[width=0.6\columnwidth]{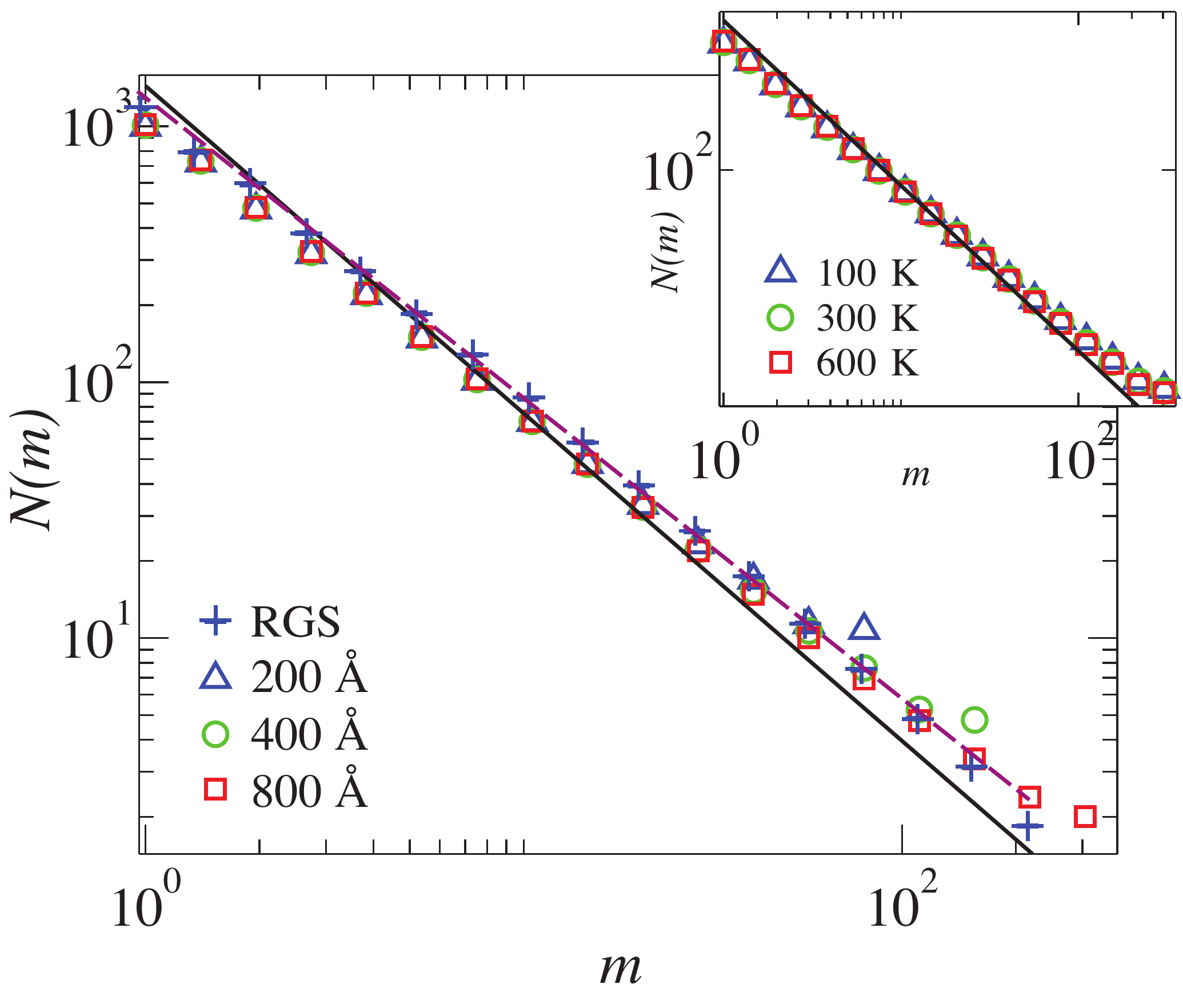}
\label{fig:fract_dim_graphene_sheet}
\caption{ Fractal dimension of iso-height lines at the percolation threshold. Main panel: Fractal dimension of the iso-height contour lines computed with the yardstick method for RGS and graphene membranes at $T=300$ K and different system sizes. Inset: Fractal dimension $d_f$ for graphene at different temperatures and system size of $800$~\AA. The solid line represents the fractal dimension $d_f = 1.28$ predicted by the SLE theory, while the dashed line corresponds to $d_f = 1.16$.}

\label{fig:FD_Graphene_and_RGS}
\end{center}
\end{figure}
We have used RGS with a mesh of size $400 \times 400$ grid points and Hurst exponent $\alpha=0.72$.   In Fig.~\ref{fig:FD_Graphene_and_RGS}, we show that the iso-height contour lines of graphene membranes and RGS have the same fractal dimension of $d_f=1.16 \pm 0.02$. By collapsing the data for different temperatures and system sizes we find that the fractal dimension is universal within error bars (see also Fig.~S3 of the supplementary material for temperatures up to $2700$~K).   
We have also measured the area enclosed by the contour lines in graphene membranes, finding that it also possesses a clear scaling exponent, i.e. $d_a=1.82 \pm 0.01$ (see Fig.~S4 of supplementary material).

\begin{figure}
\begin{center}
\includegraphics[width=0.7\columnwidth]{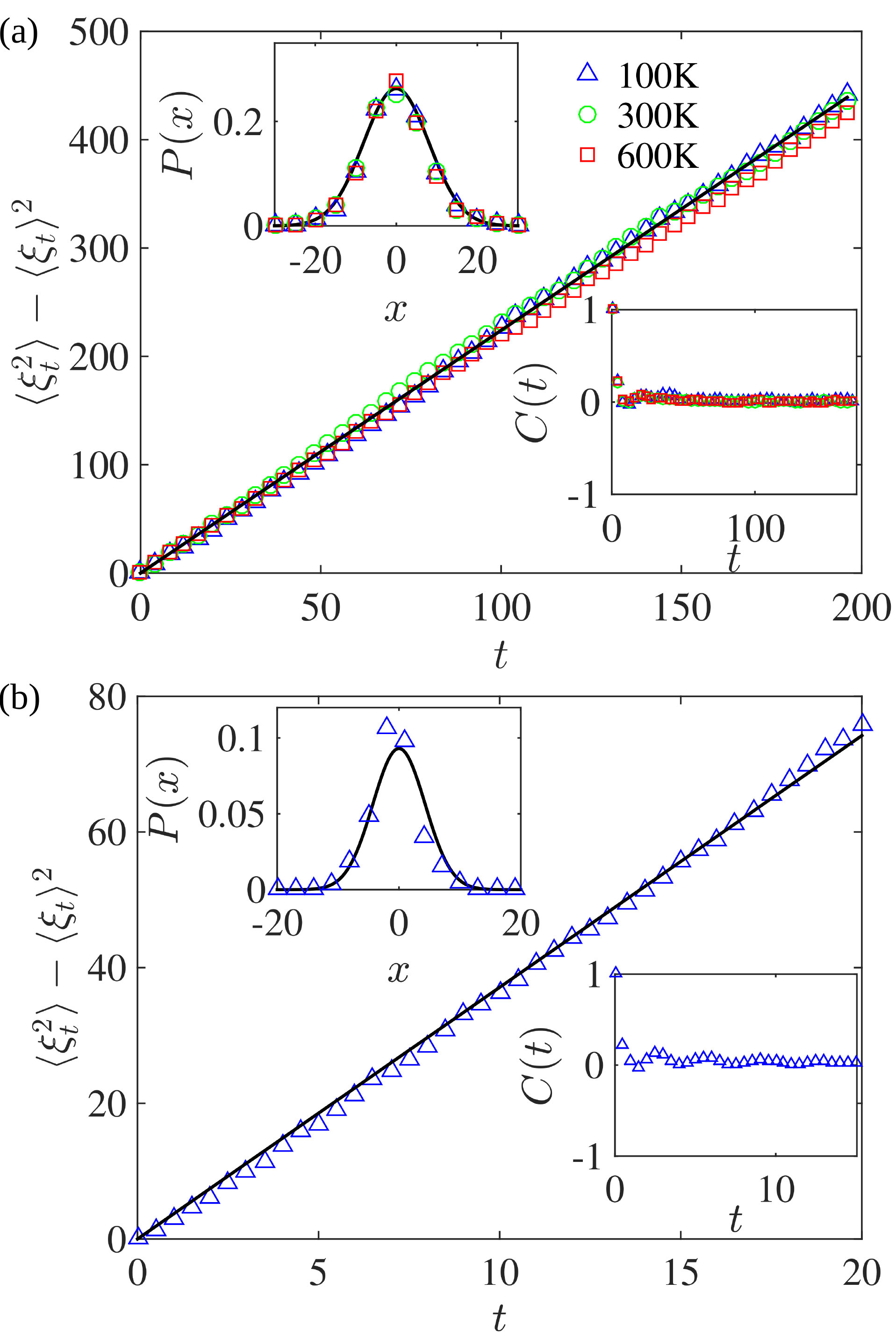}
\caption{(a)  Main panel: Variance of the driving function $\langle \xi_t^ 2\rangle$ for graphene at three different temperatures $T=100$~K, $300$~K, $600$~K. The solid line shows the linear dependence of the variance of the driving function with diffusivity $\kappa=2.24$. Upper-left inset:  The probability distribution of the driving function at $t=29$. The solid line is the probability distribution of a Gaussian random variable of zero mean and variance $2.24$. Bottom-right inset: The autocorrelation function of the increments of the driving function, averaged over the range $t=29$ to $49$.
(b) The same as in  (a), but for RGS. Here the solid line corresponds to $\kappa=3.7\pm0.8$. Upper-left inset:  The probability distribution of the driving function at $t=4.95$. The solid line is the probability distribution of a Gaussian random variable of zero mean and variance $3.7$. Bottom-right inset: The autocorrelation function of the increments of the driving function, averaged over the range $t=4.95$ to $9.95$. }
\label{fig::dSLE_graphene_sheet}
\end{center}
\end{figure}

To study conformal invariance in graphene membranes and RGS, we will use the SLE theory. In this formalism, the fractal dimension $d_f$ of SLE$_\kappa$ curves is related to the diffusion constant $\kappa$ of a Brownian walk obtained through a conformal map by \cite{Beffara08}: 
\begin{equation}
\label{eq::relation_kappa_df}
d_f=\min\left(2,1+\kappa/8\right).
\end{equation}
Therefore to test the consistency of the iso-height contour lines with SLE, one solves numerically the Loewner differential equation that describes the conformal mapping \cite{Kennedy09,Cardy05}, and compares the diffusivity $\kappa$ of the driving function with the one computed using the fractal dimension and Eq.~\eqref{eq::relation_kappa_df}. We use the so-called slit map algorithm \cite{Pose14},
to extract numerically the driving function $\xi_t$, from which we obtain the diffusivity $\kappa$. If the random curves follow SLE$_\kappa$ statistics, $\xi_t$ is a Brownian motion of variance $\kappa t$. In Fig.~\ref{fig::dSLE_graphene_sheet}~(a), we plot the variance of the driving function for graphene at three different temperatures. We find that it evolves linearly in time, and that the increments of the driving function are independent Gaussian random variables. Note that the results are independent of temperature.   The estimate of $\kappa=2.24 \pm 0.07$ predicts a fractal dimension of $d_f = 1.28 \pm 0.01$ which does not differ significantly from the fractal dimension found for graphene $d_f = 1.16 \pm 0.02$ (see Fig.~\ref{fig:FD_Graphene_and_RGS}). From this, we conclude that although we do not have strong evidence that iso-height contour line in graphene are conformally invariant our results are in fair agreement with the SLE formulation. This conclusion is strengthened through the left-passage probability test (see Fig.~S5 of the supplementary material).  
On the other hand, for the case of RGS, we obtain $\kappa = 3.7 \pm 0.8$ (see Fig.~\ref{fig::dSLE_graphene_sheet}~(b), which clearly disagrees with the diffusivity obtained from the fractal dimension using Eq.~\eqref{eq::relation_kappa_df} (the expected value for $\kappa$ is $1.46 \pm 0.16$). This provides a strong numerical evidence that the RGS are not SLE. 

\begin{figure}
\begin{center}
\includegraphics [width=0.6\columnwidth]{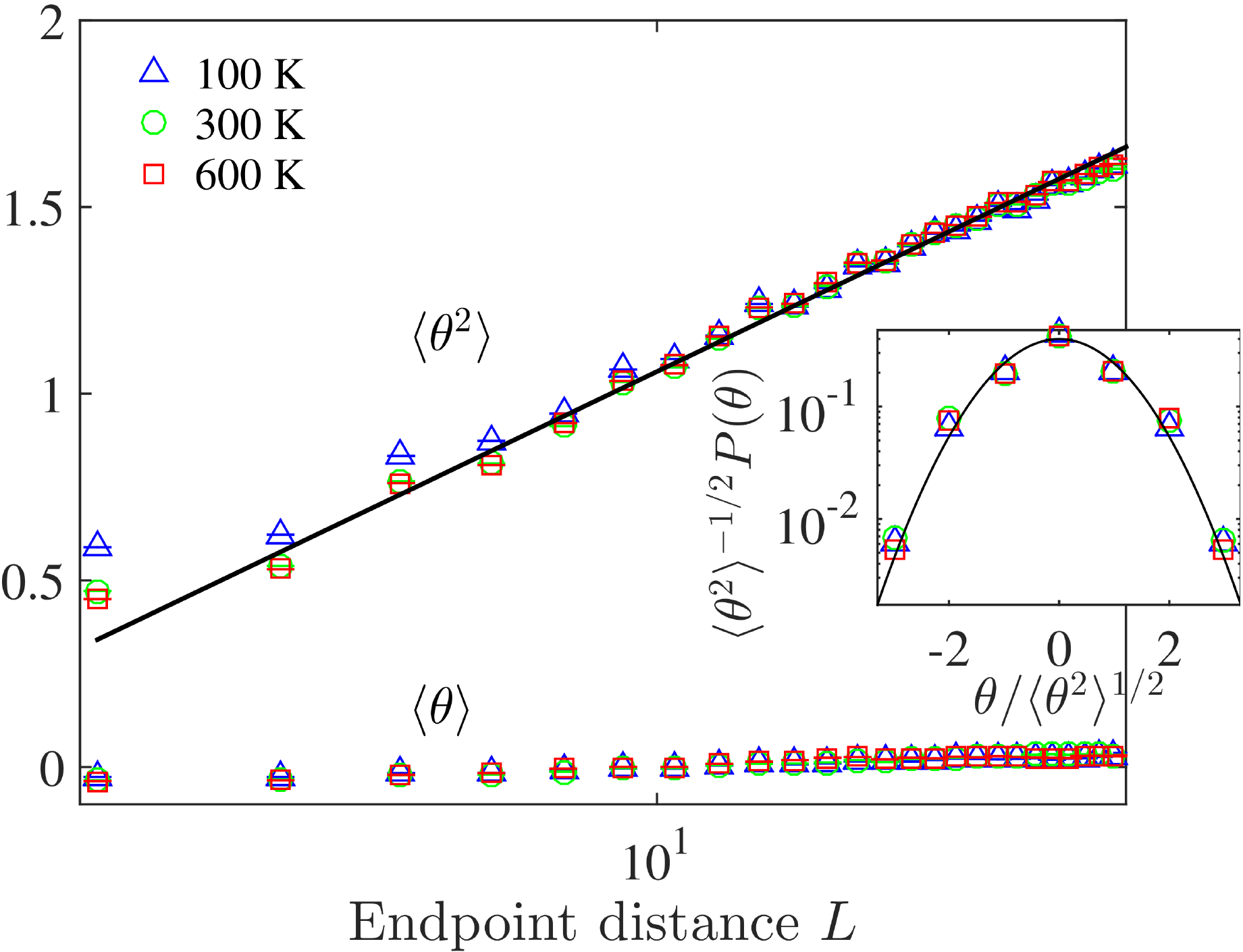}
\caption{ \label{fig::winding_angle} Main panel: Mean $\langle \theta \rangle$ and variance $\langle \theta^2 \rangle$ of the winding angle $\theta$ of the iso-height contour lines for three different temperatures $T=100$ K, $T=300$ K, and $T=600$ K, in a semi-log plot. The mean is approximately zero and the variance linear in $\ln (L)$. The solid line denotes the best fit. Inset: Rescaled probability distribution of the winding angle for $L=13.37$ \AA, compared to a  folded normal distribution with zero mean unit variance  (solid line).}  
\end{center}
\end{figure}

In order to further explore the conformal invariance of these iso-height contour lines, we study the winding angle of the curves. The winding angle $\theta$ between two points of the curve separated by a distance $L$ is defined as the oriented angle between the tangential vector of the curve at the first point and the vector connecting the two points. The winding angle distribution is expected to be Gaussian of mean $\langle \theta \rangle=0$  and variance
\begin{equation}
\label{eq::winding_angle}
\langle \theta^2 \rangle = a + 2\left(d_f-1\right)\ln L,
\end{equation}
where $d_f$ is the fractal dimension of the curves, and $a$ a constant \cite{Duplantier88, Wieland03}.   We compute the variance of the winding angle, see Fig.~\ref{fig::winding_angle}, and found a value $d_f=1.23 \pm 0.03$ for the fractal dimension of the curves using Eq.~\eqref{eq::winding_angle}, which does not differ significantly from the ones obtained with the yardstick method and SLE theory. This result is also independent of temperature. Furthermore, for a given length $L$, the distribution of the winding angle displays the expected Gaussian behaviour (see inset of Fig.~\ref{fig::winding_angle}).
The fact that the statistics of the iso-height lines satisfies Eq.~(\ref{eq::winding_angle}), gives a further numerical evidence for conformal invariance in graphene. 

\begin{figure}
\begin{center}
\includegraphics [width=0.6\columnwidth]{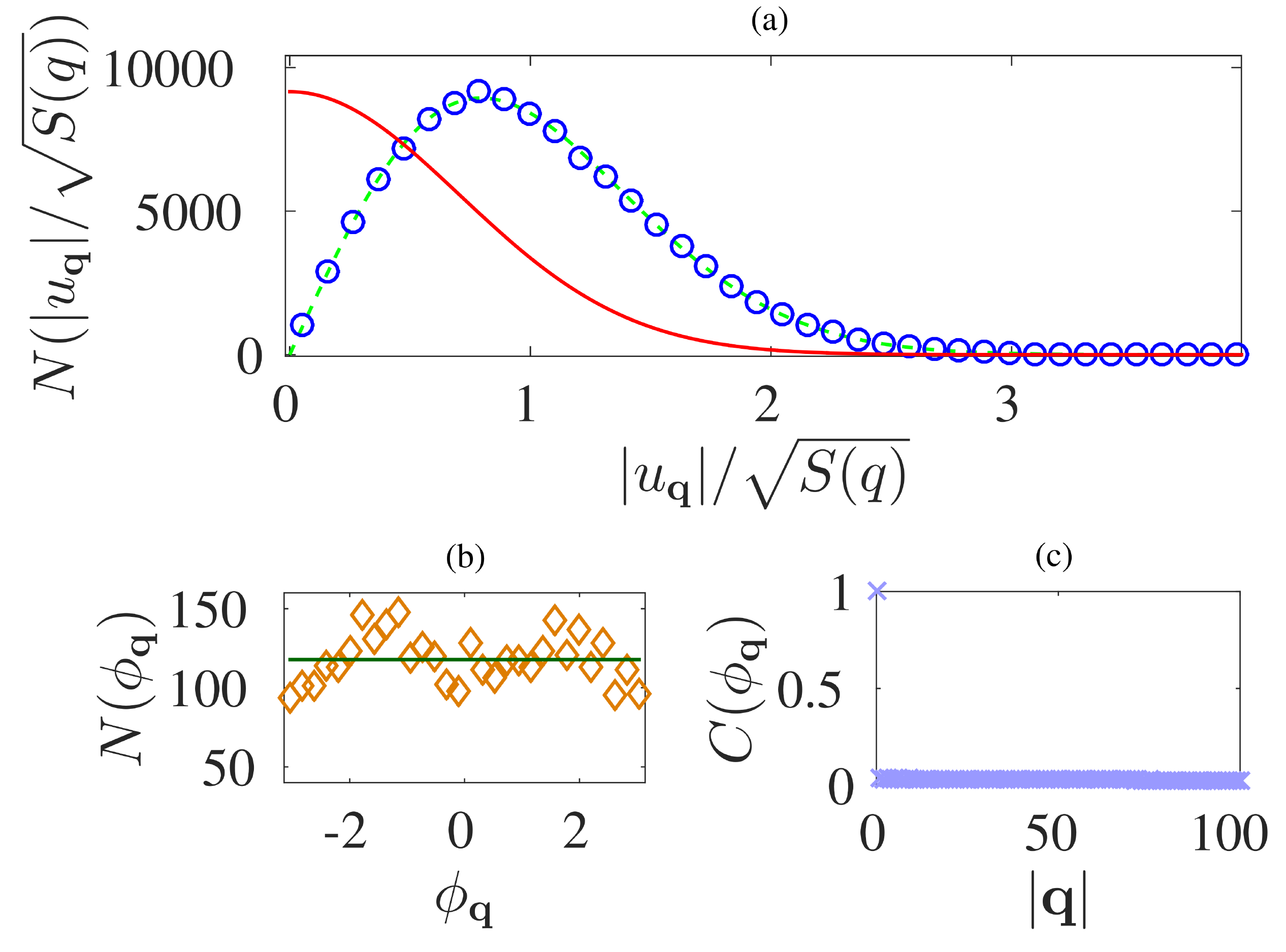}
\caption{(a) The histogram of the modulus of the Fourier coefficients for RGS and graphene. The circles represent the data obtained from the graphene samples, the green dashed line belongs to the fit performed with the function $f(|u_\mathbf{q}|)\propto |u_\mathbf{q}|\; e^{-b|u_\mathbf{q}|^2}$, while the red solid line represents the Gaussian like function that characterizes the RGS.
(b) Histogram of the phase of the Fourier coefficients at a given $\mathbf{q}$. The dark green line corresponds to a uniformed distributed phase (also the case for RGS). (c) Autocorrelation function of the phase of the Fourier coefficients $\phi_{\mathbf{q}}$ averaged over all graphene samples.  }
\label{fig:Distribution_all}
\end{center}
\end{figure}

As we have observed, not all rough self-affine surfaces satisfy SLE statistics. 
Therefore, we are interested in understanding the origin of this discrepancy. Since graphene sheets and RGS possess the same Hurst exponent, we have extended our study to Fourier space to seek for further information.
We have taken the Fourier transform of the graphene membranes and also of the RGS. By construction, the modulus of the Fourier coefficients, $\vert u_{\mathbf{q}}\vert$, of RGS at each $\mathbf{q}$ follows a Gaussian distribution  (the modulus of the coefficients follows  a folded normal distribution with zero mean unit variance). However, by performing the same measurement in graphene, we find a distribution that can be fitted with $f(|u_\mathbf{q}|) \propto |u_\mathbf{q}|\; e^{-b|u_\mathbf{q}|^2}$ (see Fig.~\ref{fig:Distribution_all}(a)). Additionally, we have also analysed the phase of the Fourier coefficients, finding that in both cases, they can be considered, with good approximation, uniformly random (see Fig.~\ref{fig:Distribution_all}(b)) and uncorrelated (for RGS, the phases are uncorrelated by construction). For measuring correlations in the phase of the graphene membranes we have computed the autocorrelation function (see Fig.~\ref{fig:Distribution_all}(c)), which falls rapidly to zero. 
Thus, one can conclude that the distribution of the modulus of the Fourier coefficients in rough self-affine surfaces might play an important role in the conformal invariance of their iso-height lines at the percolation threshold, although it does not affect the fractal dimension.  A similar connection between the distribution of the phase of the Fourier coefficients and   conformal invariance    has already been observed for contour lines of zero-vorticity in 2D turbulence \cite{bernard2006conformal}. They have shown, that by randomizing the phase, the SLE property disappears.

Summarising, we have shown that the iso-height lines of graphene sheets at the percolation threshold are scale invariant and provided evidence that they might also be conformally invariant. These properties are independent of temperature (within the range of $100$-$600$~K). Additionally, they can be described by SLE$_{\kappa}$ with $\kappa \simeq 2.24$. Conformal invariance of the iso-height lines is not a property that all rough self-affine surfaces share, as is the case of RGS. Therefore, by further analysis of graphene surfaces and RGS, we have found that the distribution of the modulus of the Fourier coefficients seems to be an important ingredient to guarantee this property.  
 The modulus of the Fourier coefficients reveals how much each mode contributes to the structure of the surface. From the distribution found in graphene sheets we can conclude that there are more active modes compared to RGS. These additional modes can change the structure and the conformal invariance property. 

This result provides insights on the long-standing question on which are the responsible properties for conformal invariance in self-affine surfaces, and can be used to construct self-affine surfaces with conformal invariance properties. Additionally, the fact that iso-height lines in graphene are conformally invariant place the study of graphene within the theory of critical phenomena, belonging to a new universality class.

For an experimental verification of our findings we need the height profile of suspended graphene membranes. However, it is still challenging to measure the height profile of suspended graphene with an accuracy in the sub-angstrom regime with current methods, because they are generally not well applicable to suspended regions of a two-dimensional crystal (due to its low binding rigidity). In the past few years, it has been possible to obtain the height profile of graphene grown on many different substrates, most commonly on SiO$_2$ \cite{geringer2009intrinsic,stolyarova2007high}. However, the substrates strongly influences the graphene membrane and thus changes its shape. 
Recently, it has been possible to measure the height profile of suspended graphene \cite{zan2012scanning}. The technique used is still at an elementary stage and has to be improved in order to obtain sub-angstrom resolution. 
However, once this method is consolidated, we will be able to observe conformal invariance in experimental samples of graphene. 

Finally, one can also explore if the scale and conformal invariance properties are typical for graphene or are present in other two-dimensional crystals. We have made tests on suspended silicene, a membrane made with silicon atoms, and unfortunately, we have found that it crumples (see Fig.~S6 of the supplementary material), preventing us to perform the same analysis. However, further crystalline structures will be analysed in the future.

\section*{Acknowledgements}
We acknowledge financial support from the European Research Council (ERC) Advanced Grant 319968-FlowCCS. 
We thank Mirko Lukovic and Caio Castro for helpful discussions.

\section*{Author Contributions}
H.H. and M.M. provided the idea and the methods used in the paper.
I.G. simulated the graphene membrane and  N.P. performed the SLE statistics.
H.H. and M.M. supervised the work.
All authors contributed to the analysis of the data, the writing process and reviewed the manuscript.

\section*{Additional Information}
Supplementary information accompanies this paper at http://www.nature.com/srep.
\newline
\textbf{Competing financial interests:} The authors declare no competing financial interests.


\end{document}